\documentclass[a4paper,11pt]{article}
\usepackage{jcappub} % for details on the use of the package, please see the JINST-author-manual
\arxivnumber{2307.01138} % Only if you have one
\title{\boldmath Uncertainty-aware and Data-efficient Cosmological Emulation using Gaussian Processes and PCA}

\author{Sven Günther}
\affiliation{Institute for Theoretical Particle Physics and Cosmology (TTK), RWTH Aachen University,\\
Sommerfeldstraße 16, 52074 Aachen, Germany}

\emailAdd{sven.guenther@rwth-aachen.de}

\abstract{Bayesian parameter inference is one of the key elements for model selection in cosmological research. However, the available inference tools require a large number of calls to simulation codes which can lead to high and sometimes even infeasible computational costs. In this work we propose a new way of emulating simulation codes for Bayesian parameter inference. In particular, this novel approach emphasizes the uncertainty-awareness of the emulator, which allows to state the emulation accuracy and ensures reliable performance. With a focus on data efficiency, we implement an active learning algorithm based on a combination of Gaussian Processes and Principal Component Analysis. We find that for an MCMC analysis of Planck and BAO data on the $\Lambda$CDM model (6 model and 21 nuisance parameters) we can reduce the number of simulation calls by a factor of $\sim$500 and save about $96\%$ of the computational costs.
}

\begin{document}
\maketitle
\flushbottom

\section{\label{sec:introduction}Introduction}

Bayesian model selection and parameter inference is a key step in the theoretical cosmology research pipeline for predicting models and comparing them with observational data. Among the most constraining observations are the two-point statistics, such as the CMB angular power spectra, that can be calculated for the investigated model by Einstein-Boltzmann solvers, such as \texttt{CAMB}\cite{camb_code} or \texttt{CLASS}\cite{class_approximations}. In practice, this is often done through statistical model inference runs that depend on a multitude of calls of these solvers, which can result in high computational costs that may make parameter inference unfeasible. To overcome this issue, a variety of approaches have been developed in order to emulate Einstein-Boltzmann solvers and reduce the computational cost. One can distinguish between emulators that are trained prior to the inference run\cite{G_nther_2022}\cite{cosmopower}\cite{pico_code} and emulators with active learning that draw samples in the most relevant region of the parameter space and get trained during the inference run\cite{nygaard2022connect}. In this work we propose an active learning emulator that combines the data compression of Principal Component Analysis and the data efficiency and uncertainty prediction of Gaussian Processes. 

\section{\label{sec:theory}Theory}

\subsection*{\label{sec:gp}Gaussian Processes}
Gaussian Processes (GPs) are the generalisation of Gaussian probability distributions to functions. They can be formulated as a regression task to fit a sufficiently smooth function $f(\mathbf{x})$ given a set of data points $X$. They assume a Gaussian probability distribution for the function at each evaluation point and provide an estimate for both the mean $\hat{\mu}$ and the covariance matrix $\hat{\Sigma}$ (later denoted as $\hat{\sigma}^2$ for a single dimension) at any point $\mathrm{x}$ conditioned by $X$,

$$\hat{\mu}(\mathbf{x}),\hat{\Sigma}(\mathbf{x}) = \mathcal{GP}(\mathbf{x}|X).$$

In our emulator, we assume that the training data has no internal stochastic component. As a result, the predicted covariance originates only from the sampling sparsity of the data set and the smoothness of the emulated function. The degree of smoothness is encoded in the \textit{kernel}, which is a parametrization of the correlation between two samples $\mathbf{x},\mathbf{x'}$ following the function $f$. For this emulator we use \textit{anisotropic radial basis functions} (RBF) for the kernel 
$$k(\mathbf{x},\mathbf{x'}) = \sigma^2 \mathrm{exp} \left( - \sum \frac{(\mathbf{x}_i-\mathbf{x'}_i)^2}{2 \mathbf{L}_i^2} \right). $$
The parameters $\sigma,\mathbf{L}$ are fitted to the data. A detailed summary of GPs can be found in \cite{rasmussen2005gaussian} or \cite{10.5555/1162264}. We use the GP implementation of the python \texttt{scikit-learn} library \cite{scikit-learn}. GPs have been used in the cosmological community for example in \cite{cosmopower} and \cite{gammal2022fast,Pellejero_Iba_ez_2020}. In the latter references, the authors confirmed the data efficiency of GPs.

\subsection*{\label{sec:pca}Principal Component Analaysis}

Observables in cosmological research, such as matter power spectra or the CMB, are of high dimension $\mathcal{O}(10^3)$ and strongly correlated. Predicting each component results in a high computational effort. As a consequence, we find that it is advantageous to perform a data compression to a lower dimensional feature space. In this work we use \textit{Prinicpal Component Analysis} (PCA) which transforms a data set $X=(X_1,...,X_N)$ with $X_i\in \mathbb{R}^D$ into the eigenspace of the matrix $X^T X$ by solving

$$\vec{w}_i = \lambda_i (X^T X) \vec{w}_i .$$

The eigenvectors $(\vec{w}_1,..,\vec{w}_M)$ with the $M$ largest eigenvalues form an orthogonal PCA-eigenspace. The eigenvectors provide the transformation matrix of dimension $D \times M$. In general, it is not invertible, such that a loss of information occurs. We find that using $\sim 10-15$ PCA components on the example of CMB spectra (see in \ref{sec:example}) maintains a loss of precision which proves to be sufficiently accurate for parameter inference. The two main advantages are dimensionality reduction as well as orthogonality, since each PCA component can be emulated independently of another. This allows for fast and parallelizable fitting and evaluation of the GP. PCAs have been used in the context of cosmological model inference for example in \cite{dsh_code}. Further reading can be found in \cite{10.5555/1162264}.

\subsection*{\label{sec:uncertainty}Uncertainty qualification}

For the uncertainty qualification of the emulator prediction, we consider two sources of noise. The first is the loss of information we encounter by using the PCA. We estimate it by transforming a set of spectra into the PCA representation and untransforming it back into the data space. The residuals of the original spectra with their PCA versions contain the information loss. The corresponding uncertainty estimate will be the standard deviation in each dimension in the data space (e.g. in each $\ell$ bin). When sampling from this noise, we correlate the errors of each bin. We assume this uncertainty estimate to be constant for the given parameter space such that it provides a maximal achievable accuracy for a given number of PCA components. By increasing this number, the maximal precision of emulation can be enhanced.

The second source is related to the uncertainty estimate of the GP which reflects the sampling density. The uncertainty in the emulated quantity can be inferred from
the individual uncertainty estimates of the GP prediction. Accordingly, we transform the uncertainty estimates of the GP for the PCA components $\hat{\sigma}_i$ for $i=1,..,M$ into the data space by multiplying with the PCA transformation matrix. Due to the orthogonality of the PCA components we can derive the uncertainty estimates of the PCA components independently of each other, so that we can propagate those uncertainties to the total uncertainty of the data

$$\hat{\vec{\sigma}}_x = \sqrt{\sum_{i=1}^{M} \vec{w}_i \odot \vec{w}_i   \hat{\sigma}_i^2}.$$

\begin{figure}[h]
\centering
\includegraphics[width=12.5cm]{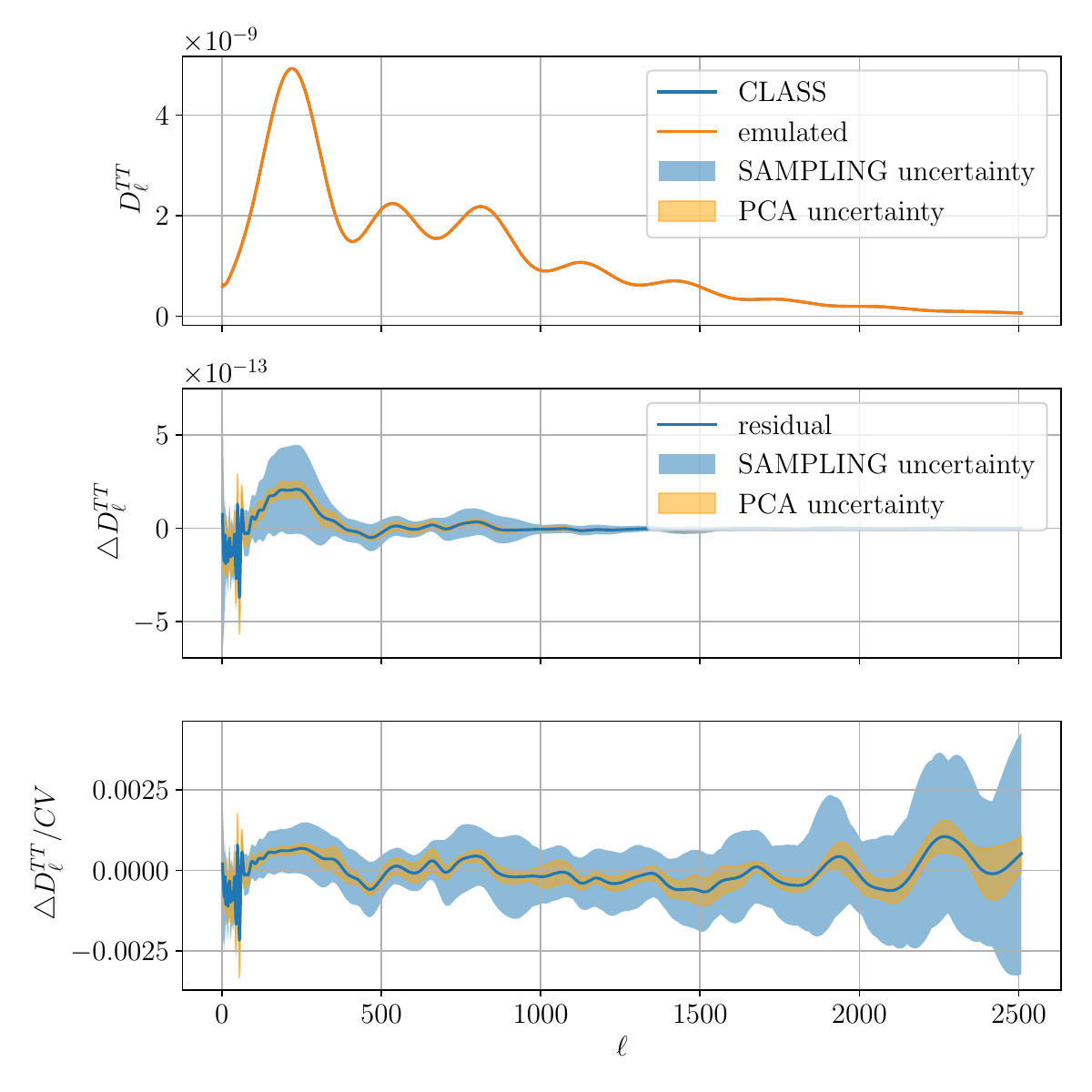}% Here is how to import EPS art
\caption{\label{fig:unc} Uncertainty qualification of the CMB TT spectrum emulator for a training scenario similar to the one outlined in section \ref{sec:implementation}. We compare the performance of the emulator with the full calculation obtained by \texttt{CLASS}. (Top) $D_\ell$ spectrum. (Center) Residuals with the uncertainty estimates related to the PCA information loss and the sampling sparsity of the GPs. (Bottom) Residuals normalized by the cosmic variance.}
\end{figure}

Thus, the uncertainty estimates on the uncorrelated PCA components mix into a correlated error on the data, reflecting the mixing of the different emulator components. Unlike the first source of uncertainty, this one can be reduced by training the GP on more data, which allows for a more accurate prediction of the PCA components. An emulator call of the scaled CMB temperature power spectrum $D_\ell^\mathrm{TT}$ with the uncertainty contributions is displayed in figure \ref{fig:unc}.

\subsection*{\label{sec:online_learning}Online Learning Strategy}
Online learning (see \cite{MohriRostamizadehTalwalkar18}) is a training technique in which data become available in a sequential order. This sequential order can be, for example, the steps of a sampling algorithm. Before each new step, an optimal predictor (given the available data from the previous steps) is constructed. We combine this learning strategy with the concept of active learning (see \cite{MohriRostamizadehTalwalkar18}), which assigns an importance to each new data point for the training of the emulator. The importance criterion we use for active learning is the uncertainty estimate of the emulator. Since the significance of an uncertainty can be interpreted in terms of likelihood, we qualify the uncertainty of the emulator by drawing a set of samples from the uncertainty estimate of the emulator. For each of the samples, we evaluate the likelihood of the corresponding observation and calculate the mean and standard deviation of the set. Consequently, the standard deviation (and mean) can be used as an estimator of the emulator's uncertainty on the prediction. If the estimator falls below a user defined accuracy criterion for the likelihood, we mark that data point as relevant for training the emulator. The corresponding observable is then calculated by the theory code and added to the training set. Accordingly, the emulator is trained on the full data set to allow a more accurate prediction on the next sample. However, if the estimator exceeds the criterion, the prediction from the emulator is used. This ensures a user defined and easily interpretable accuracy of the emulator (for uncertainty estimates on the posterior see \cite{Grand_n_2022}). The accuracy estimator can be specified by the user. We find that a sum of a constant offset and a likelihood dependent part ensures performance in the high likelihood region and relaxes the criterion for more implausible observations. We use the emulator's prediction when the estimated standard deviation of the log likelihood $\hat{\sigma}_{\log \mathcal{L}}$ satisfies
$$\hat{\sigma}_{\log \mathcal{L}} < 0.1 + 0.1 \cdot (\log \mathcal{L}_\mathrm{best fit} - \log \hat{\mathcal{L}} ).$$

\section{\label{sec:implementation}Implementation}
We implement our emulator approach in a modified version of the Bayesian sampler \texttt{cobaya} \cite{cobaya} in a model and data independent way. While single or low dimensional quantities (such as $\sigma_8$ or redshift-binned angular diameter) are emulated using GPs, the high dimensional observables such as the CMB spectra are emulated with the combination of PCA and GP as outlined above. The number of PCA components is adjustable, but is currently set to values between $10-20$ for the cosmologically relevant spectra. We find that starting the training of the emulator immediately proves not to be efficient, as early samples often consist of irrelevant outliers (for example burn-in of MCMCs). To circumvent this issue, we remove all outlier points associated with loglikelihood values that are too far away from the best-fit point (at the current stage of the chain). In addition, we have implemented a minimum number of samples required to start the training, as a small number of samples will not result in generalizable PCA components. Whenever the theory code is called, the emulator generates its prediction with a number of noisy samples $\sim  (3-10)$. These noisy samples are then compared with the observations in the likelihood codes. When a new data point is added to the emulator, a new GP is computed that takes the additional data point into account. The kernels are fitted only in the beginning of training or every $20$ new data points to save computational cost.

The current code implementation, which includes an explanation of all parameters, can be found in a forked {cobaya} version at \texttt{svenguenther/cobaya}.

\section{\label{sec:example}Example}

\begin{table}
\centering
\caption{\label{tab:bias}Biases of selection of posterior estimates}
%%\begin{ruledtabular}
\begin{tabular}{cccc}
Parameter& Mean Emulator & Mean Default & Bias\\
\hline
$h$ & 0.6804(45) & 0.6804(45) & 0.004\\
$\log(10^{10} A_\mathrm{s})$ & 3.043(16) & 3.042(16) & 0.05\\
$n_\mathrm{s}$ & 0.9654(38) & 0.9653(38) & 0.03\\
$\Omega_\mathrm{b}h^2$ & 0.02238(14) & 0.02238(14) & 0.01\\
$\Omega_\mathrm{cdm}h^2$ & 0.1196(10) & 0.1196(10) & -0.01\\
$\tau_\mathrm{reio}$ & 0.0542(77) & 0.538(76) & 0.05\\
$y_\mathrm{cal}$ & 1.0006(25) & 1.0006(25) & 0.007\\

\end{tabular}
%%\end{ruledtabular}
\end{table}

\begin{figure*}
\includegraphics[width=15.5cm]{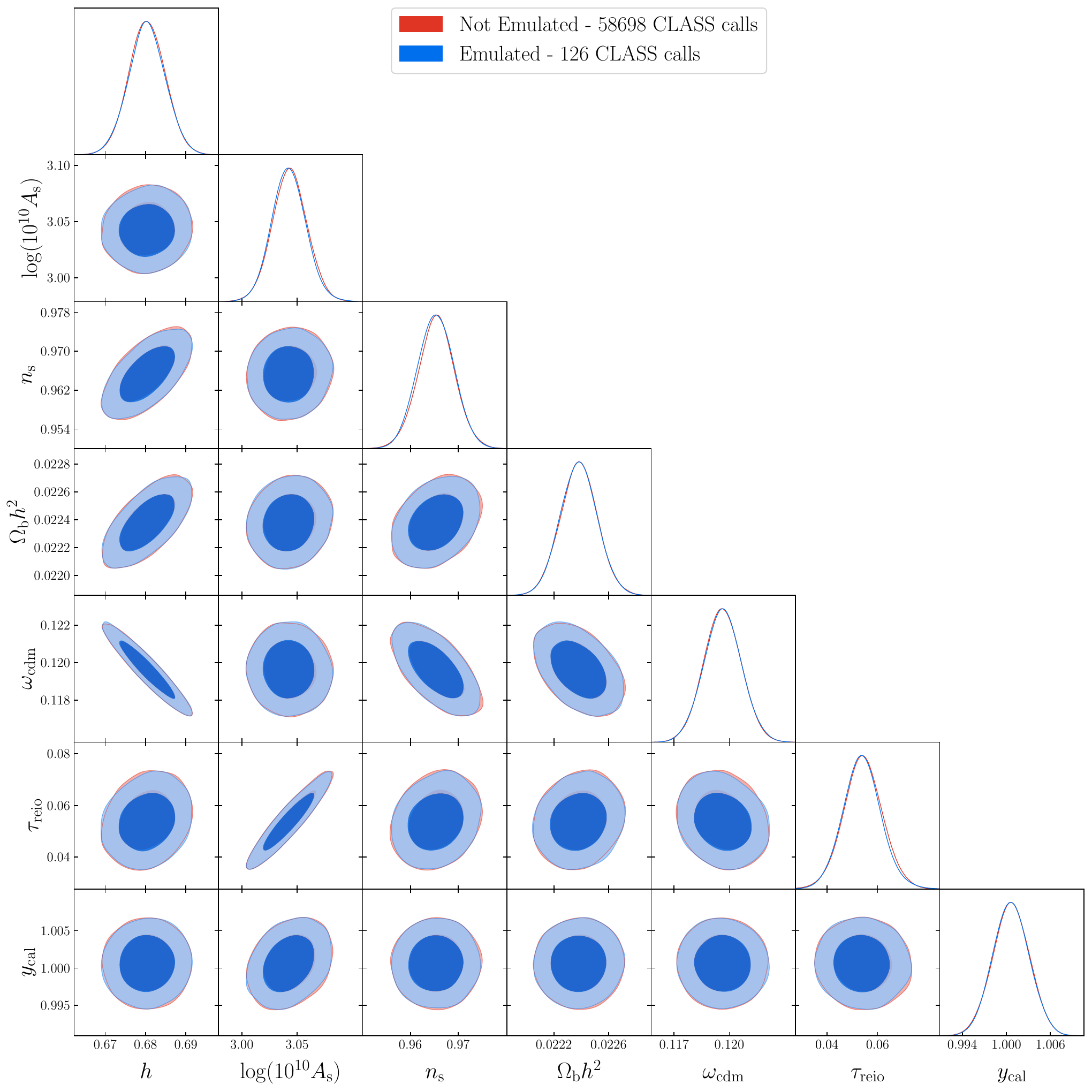}% Here is how to import EPS art
\caption{\label{fig:mcmc} Predicted posterior estimate for the $\Lambda \mathrm{CDM}$ model tested on BAO and Planck TT,TE and EE data. (Blue) posterior estimate using the emulator. It was trained with 126 calls of the theory code. (Red) Comparative MCMC without the use of the emulator. The contours were generated with 58698 theory calls.}
\end{figure*}

In the test case, we test the 6 parameter $\Lambda \mathrm{CDM}$ model on the Planck 2018 data release \cite{2020} including low and high $\ell$ temperature and polarization spectra and SDSS-III BAO measurements from DR12 \cite{Alam_2017}. We assess the performance of the outlined emulator concept by running an MCMC with and without the emulator. In order to utilize the speed hierarchy of fast nuisance parameters (21 of the Planck likelihood) and slow cosmological parameters, we use the concept of oversampling \cite{Lewis_2013} and run the chains until the $R-1$ convergence criterion \cite{10.1214/ss/1177011136} reaches $0.01$. We find good agreement between the posterior estimates, which are shown in figure \ref{fig:mcmc}. We compute the bias metric $b$ with the mean $m$ and standard deviation $\sigma$ of the posterior estimates,
$$b = (m_\mathrm{default}-m_\mathrm{emulator})/ \sigma_\mathrm{default} .$$
The means and biases for the 6 cosmological parameters and the Planck calibration are at most 5\%, as displayed in Table \ref{tab:bias}. We find that 126 calls of the theory code were required during the convergence of the MCMC to train the emulator. A substantial fraction of these calls are at the beginning of the MCMC and can be associated with the burn-in phase. In contrast, the comparative MCMC required 58924 theory calls, which dominate the total run time of 180 CPUh (1 chain with 4 threads for a wall time of 45 hours). When using the emulator, the limiting factors for speedup are the calls of the likelihood code used for the precision estimation of the emulator. The total run time for both training the emulator and converging the MCMC is about 7 CPUh (1 chain with 1 thread). We find similar speedup when running MCMCs without an initial guess for the covariance matrix. However, the number of theory calls did increase slightly. We also find that this number depends on the starting point of the chain, but has only a small effect on the overall run time.

\section{Discussion}

In this work we have implemented an emulator with a data efficient online learning algorithm that is capable of providing an uncertainty estimate for its prediction. This allows us to use an active learning algorithm that ensures a reliable performance of the emulator with an interpretable uncertainty estimate on the likelihood. The combination of active and online learning is complemented by an architecture of PCAs and GPs. We find that this model- and data-independent approach is able to reduce the computational costs by 96\% for the test case of the $\Lambda \mathrm{CDM}$ model on Planck and BAO data. The number of theory computations could be reduced by a factor of $\sim 500$ to 126 computation calls. In further work we plan to optimize the algorithm, perform a major code release, release a \texttt{MontePython} \cite{brinckmann2018montepython} version, explore the hyperparameters, and further improve the performance by parallelizing the emulator. We estimate that the speedup would be even larger with more computationally expensive cosmological models, which are even more dominated by the calculation time of the theory code. Finally, we highlight the potential to exploit the differentiability of this approach to further speed up Bayesian parameter inference with gradient based sampling algorithms \cite{DUANE1987216}.

\acknowledgments

I acknowledge support from the DFG grand LE 3742/6-1. Furthermore, I would like to thank my collaborators Johanna Schafmeister and Santiago Casas for very fruitful discussions leading to the idea of this paper. In particular, I would like to thank the Aarhus group of Thomas Tram, Steen Hannestad, Emil Holm and Andreas Nygaard (+ all Master students) for many discussions on a variety of topics and a great stay in their group. Furthermore, I would like to thank Nils Schöneberg for providing valuable comments. Last but not least, I would like to thank Julien Lesgourgues and Felicitas Keil for proofreading and giving valuable advice. 

% Bibliography

%% [A] Recommended: using JHEP.bst file
%% \bibliographystyle{JHEP}
%% \bibliography{biblio.bib}

%% or
%% [B] Manual formatting (see below)
%% (i) We suggest to always provide author, title and journal data or doi:
%% in short all the informations that clearly identify a document.
%% (ii) please avoid comments such as "For a review'', "For some examples",
%% "and references therein" or move them in the text. In general, please leave only references in the bibliography and move all
%% accessory text in footnotes.
%% (iii) Also, please have only one work for each \bibitem.

\bibliographystyle{JHEP}
\bibliography{biblio.bib}

\providecommand{\href}[2]{#2}\begingroup\raggedright\begin{thebibliography}{10}

\bibitem{camb_code}
A.~Lewis and A.~Challinor, ``Camb: Code for anisotropies in the microwave background.'' Astrophysics Source Code Library.

\bibitem{class_approximations}
D.~Blas, J.~Lesgourgues and T.~Tram, \emph{The cosmic linear anisotropy solving system (class). part ii: Approximation schemes}, \href{https://doi.org/10.1088/1475-7516/2011/07/034}{\emph{Journal of Cosmology and Astroparticle Physics} {\bfseries 2011} (2011) 034–034}.

\bibitem{G_nther_2022}
S.~Günther, J.~Lesgourgues, G.~Samaras, N.~Schöneberg, F.~Stadtmann, C.~Fidler et~al., \emph{{CosmicNet} {II}: emulating extended cosmologies with efficient and accurate neural networks}, \href{https://doi.org/10.1088/1475-7516/2022/11/035}{\emph{Journal of Cosmology and Astroparticle Physics} {\bfseries 2022} (2022) 035}.

\bibitem{cosmopower}
A.S.~Mancini, D.~Piras, J.~Alsing, B.~Joachimi and M.P.~Hobson, \emph{Cosmopower: emulating cosmological power spectra for accelerated bayesian inference from next-generation surveys},  2021.

\bibitem{pico_code}
B.~Fendt, Chad:~Wandelt, ``Pico: Parameters for the impatient cosmologist.'' Astrophysics Source Code Library.

\bibitem{nygaard2022connect}
A.~Nygaard, E.B.~Holm, S.~Hannestad and T.~Tram, \emph{Connect: A neural network based framework for emulating cosmological observables and cosmological parameter inference},  2022.

\bibitem{rasmussen2005gaussian}
C.E.~Rasmussen and C.K.I.~Williams, \emph{Gaussian Processes for Machine Learning (Adaptive Computation and Machine Learning)}, The MIT Press (2005).

\bibitem{10.5555/1162264}
C.M.~Bishop, \emph{Pattern Recognition and Machine Learning (Information Science and Statistics)}, Springer-Verlag, Berlin, Heidelberg (2006).

\bibitem{scikit-learn}
F.~Pedregosa, G.~Varoquaux, A.~Gramfort, V.~Michel, B.~Thirion, O.~Grisel et~al., \emph{Scikit-learn: Machine learning in {P}ython}, {\emph{Journal of Machine Learning Research} {\bfseries 12} (2011) 2825}.

\bibitem{gammal2022fast}
J.E.~Gammal, N.~Schöneberg, J.~Torrado and C.~Fidler, \emph{Fast and robust bayesian inference using gaussian processes with gpry},  2022.

\bibitem{Pellejero_Iba_ez_2020}
M.~Pellejero-Ibañez, R.E.~Angulo, G.~Aricó, M.~Zennaro, S.~Contreras and J.~Stücker, \emph{Cosmological parameter estimation via iterative emulation of likelihoods}, \href{https://doi.org/10.1093/mnras/staa3075}{\emph{Monthly Notices of the Royal Astronomical Society} {\bfseries 499} (2020) 5257–5268}.

\bibitem{dsh_code}
M.~Kaplinghat, L.~Knox and C.~Skordis, \emph{{Rapid calculation of theoretical cmb angular power spectra}}, \href{https://doi.org/10.1086/342656}{\emph{Astrophys. J.} {\bfseries 578} (2002) 665} [\href{https://arxiv.org/abs/astro-ph/0203413}{{\ttfamily astro-ph/0203413}}].

\bibitem{MohriRostamizadehTalwalkar18}
M.~Mohri, A.~Rostamizadeh and A.~Talwalkar, \emph{Foundations of Machine Learning}, Adaptive Computation and Machine Learning, MIT Press, Cambridge, MA, 2~ed. (2018).

\bibitem{Grand_n_2022}
D.~Grandón and E.~Sellentin, \emph{Differentiable predictions for large scale structure with shamnet}, \href{https://doi.org/10.21105/astro.2205.11587}{\emph{The Open Journal of Astrophysics} {\bfseries 5} (2022) }.

\bibitem{cobaya}
J.~Torrado and A.~Lewis, \emph{Cobaya: code for bayesian analysis of hierarchical physical models}, \href{https://doi.org/10.1088/1475-7516/2021/05/057}{\emph{Journal of Cosmology and Astroparticle Physics} {\bfseries 2021} (2021) 057}.

\bibitem{2020}
P.~Collabo10.1051/0004-6361/201936386ration, \emph{Planck 2018 results}, \href{https://doi.org/10.1051/0004-6361/201936386}{\emph{Astronomy {\&} Astrophysics} {\bfseries 641} (2020) A5}.

\bibitem{Alam_2017}
S.-I.~Collaboration, \emph{The clustering of galaxies in the completed {SDSS}-{III} baryon oscillation spectroscopic survey: cosmological analysis of the {DR}12 galaxy sample}, \href{https://doi.org/10.1093/mnras/stx721}{\emph{Monthly Notices of the Royal Astronomical Society} {\bfseries 470} (2017) 2617}.

\bibitem{Lewis_2013}
A.~Lewis, \emph{Efficient sampling of fast and slow cosmological parameters}, \href{https://doi.org/10.1103/physrevd.87.103529}{\emph{Physical Review D} {\bfseries 87} (2013) }.

\bibitem{10.1214/ss/1177011136}
A.~Gelman and D.B.~Rubin, \emph{{Inference from Iterative Simulation Using Multiple Sequences}}, \href{https://doi.org/10.1214/ss/1177011136}{\emph{Statistical Science} {\bfseries 7} (1992) 457 }.

\bibitem{brinckmann2018montepython}
T.~Brinckmann and J.~Lesgourgues, \emph{Montepython 3: boosted mcmc sampler and other features},  2018.

\bibitem{DUANE1987216}
S.~Duane, A.~Kennedy, B.J.~Pendleton and D.~Roweth, \emph{Hybrid monte carlo}, \href{https://doi.org/https://doi.org/10.1016/0370-2693(87)91197-X}{\emph{Physics Letters B} {\bfseries 195} (1987) 216}.

\end{thebibliography}\endgroup

\end{document}